\documentclass[conference]{IEEEtran}
\IEEEoverridecommandlockouts

\usepackage{cite}
\usepackage{amsmath,amssymb,amsfonts}
\usepackage{graphicx}
\usepackage{textcomp}
\usepackage{xcolor}
\usepackage[compact]{titlesec}
\usepackage{caption}

\usepackage{color}
\usepackage{multirow}
\usepackage{algorithm}
\usepackage[noend]{algpseudocode}
\usepackage[export]{adjustbox}

\usepackage{setspace}
\usepackage[compact]{titlesec}

\usepackage[marginal]{footmisc}

\usepackage[a4paper, total={184mm,239mm}]{geometry}
\def\BibTeX{{\rm B\kern-.05em{\sc i\kern-.025em b}\kern-.08em
    T\kern-.1667em\lower.7ex\hbox{E}\kern-.125emX}}
\DeclareRobustCommand*{\IEEEauthorrefmark}[1]{%
    \raisebox{0pt}[0pt][0pt]{\textsuperscript{\footnotesize\ensuremath{#1}}}}
    
\begin{document}
\title{HAAN: A Holistic Approach for Accelerating Normalization Operations in Large Language Models}

\author{
    \IEEEauthorblockN{
        Tianfan Peng\IEEEauthorrefmark{1,2}, 
        Jiajun Qin\IEEEauthorrefmark{1,3}, 
        Tianhua Xia\IEEEauthorrefmark{1}, 
        Sai Qian Zhang\IEEEauthorrefmark{1}
    }
    \IEEEauthorblockA{
        \IEEEauthorrefmark{1}Tandon School of Engineering, New York University, New York, USA \\
        \IEEEauthorrefmark{2}Shenzhen Institute of Information Technology, Shenzhen, China \\
        \IEEEauthorrefmark{3}Zhejiang University, Hangzhou, China \\
        Email: 
            \{tianfanpeng, hobbitqia\}@gmail.com,
            \{tx856, sai.zhang\}@nyu.edu
    }
}

\maketitle

\begin{abstract}
Large language models (LLMs) have revolutionized natural language processing (NLP) tasks by achieving state-of-the-art performance across a range of benchmarks. Central to the success of these models is the integration of sophisticated architectural components aimed at improving training stability, convergence speed, and generalization capabilities. Among these components, normalization operation, such as layer normalization (LayerNorm), emerges as a pivotal technique, offering substantial benefits to the overall model performance. However, previous studies have indicated that normalization operations can substantially elevate processing latency and energy usage. In this work, we adopt the principles of algorithm and hardware co-design, introducing a holistic normalization accelerating method named~\textit{HAAN}. The evaluation results demonstrate that HAAN can achieve significantly better hardware performance compared to state-of-the-art solutions.
\end{abstract}

\section{Introduction}

The superior performance of large language models (LLMs) is accompanied by increased computational and memory consumption.
The widespread adoption of LLMs has pushed transformer sizes to unprecedented scales, often reaching multiple billions and continuing to grow. Besides the growing model size, LLM computation involves a blend of linear matrix multiplication and non-linear operations, including normalization, softmax, and GeLU. In contrast to other DNN architectures such as Convolutional Neural Networks (CNNs), where matrix multiplication dominates, transformers dedicate a considerable portion of their runtime to nonlinear operations~\cite{khan2021npe,hong2022dfx,wang2021spatten}.

Among the numerous normalization operations proposed, Layer Normalization (LayerNorm)~\cite{ba2016layer} and Root Mean Square Normalization (RMSNorm)~\cite{zhang2019root} are two most widely employed normalization techniques in LLM. LayerNorm adjusts the input vectors of hidden layers by re-centering and rescaling them, thereby producing outputs with zero mean and unit variance. This process mitigates internal covariate shift issues and greatly enhances the training efficiency~\cite{xu2019understanding}. Conversely, RMSNorm adopts a more efficient approach by solely rescaling the input vector using its root mean square (RMS) value, offering superior computational efficiency compared to LayerNorm.

Non-linear operations in transformers, particularly LayerNorm and softmax, are notoriously time-consuming~\cite{wang2023sole,yu2022nn, khan2021npe,hong2022dfx}. The recent surge in popularity of FlashAttention~\cite{dao2022flashattention} has reduced softmax latency by $80\%$ by integrating it with matrix multiplication, and several hardware accelerators~\cite{nuaa,zfw2018,softermax,base2sm,xia2023softmax} have further optimized softmax. Additionally, multiple methods have been proposed to facilitate matrix multiplication operations within LLMs by representing them with extremely low precision (e.g., FP8, INT8)~\cite{quip, quarot, frantar2022optimal, frantar2023gptq}. However, efficient solutions for LayerNorm are still limited.

\begin{figure}
    \centering
    \includegraphics[width=\columnwidth]{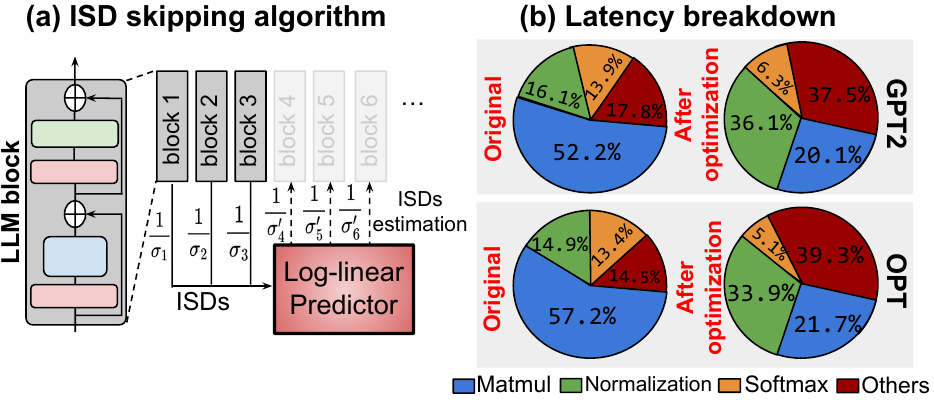}
    \caption{(a) Standard deviation skipping for efficient normalization. (b) Runtime breakdown for GPT-2 and OPT execution with and without applying optimization techniques, using a sequence length of 2048.}
    \label{fig:haan_algorithm}
    \vspace{-6mm}
\end{figure}

To demonstrate the cost of normalization operations in LLM execution, we profile the GPT2-117M and OPT-2.7B models from Huggingface~\cite{wolf2020huggingfaces} using half-precision (FP16) on an A100 GPU. As illustrated in the latency breakdown in Figure~\ref{fig:haan_algorithm}(b), LayerNorm computation constitutes a substantial portion of model runtime (approximately 16\% in the GPT2 model). Additionally, we evaluate the latency of LayerNorm by applying optimization techniques such as FlashAttention and FP8 quantization for linear layers within the LLM. The results indicate that, while these optimizations greatly reduce the processing latency of softmax and Matmul operations, LayerNorm becomes a critical bottleneck, contributing to more than $33\%$ of the overall latency. Thus, speeding up normalization is essential.
Accelerating the normalization operation poses a considerable challenge, primarily due to the complexity of hardware implementation. This complexity stems from two main factors. Firstly, the square root and division operations required for variance computation are computationally intensive and resource-demanding. Secondly, the data dependencies in normalization hinder effective pipelining and parallelism, causing high latency and energy consumption.

In this paper, we introduce~\textit{HAAN}, a holistic approach for accelerating normalization operation within LLM. In particular, our contribution can be summarized as follows:
\begin{itemize}
    \item Our analysis reveals a strong correlation in the input statistics
across consecutive attention layers within LLMs. Leveraging
this insight, we introduce an efficient technique to
estimate normalization statistics by utilizing the statistics
of early LLM layers (Figure 1). This approach substantially
mitigates the computational cost of the normalization
operation with negligible accuracy degradation.
    \item We demonstrate that subsampling the LayerNorm input for computing input statistics leads to significant reductions in latency and power consumption during hardware implementation, with minimal impact on accuracy.
    \item The HAAN accelerator uses a hybrid of fixed and floating-point formats to streamline normalization computations, while offering configurability for selecting the optimal numeric format and input subsampling rate for LayerNorm. 
    \item Evaluation results show that HAAN reduces power consumption by over $60\%$ and latency by $20\%$ compared to baseline methods across various LLM workloads, while maintaining superior accuracy.
\end{itemize}

\section{Background and Related Work}
\label{sec:background}
\subsection{Normalization Operation within LLM}
\label{sec:layernorm}

Normalization operation has been widely utilized in the modern LLM architecture. Normalization can help in faster convergence during training by reducing the internal covariate shift problem~\cite{ba2016layer}. It ensures that the distributions of inputs to each layer remain stable throughout training. Two common normalization operations are widely employed: LayerNorm and RMSNorm~\cite{zhang2019root}. RMSNorm is utilized in LLMs such as the LLaMA series~\cite{touvron2023llama, touvron2023llama2} and Mistral~\cite{jiang2023mistral}, while LayerNorm is employed in LLMs like OPT~\cite{zhang2022opt}, the GPT series~\cite{brown2020language}, and Megatron-LM~\cite{shoeybi2019megatron}.
Specifically, for layer normalization function, the function takes a vector $\boldsymbol{z}=[z_1,\cdots,z_N]^T$ and generates an output $\boldsymbol{s}=[s_1,\cdots,s_N]^T$, both have a length of N. It is defined as follows:
\begin{equation}
\label{eqn:layernorm}
\small
\boldsymbol{s}=\alpha\frac{\boldsymbol{z}-\mu_{z}}{\sigma_{z}} + \beta
\end{equation}
where $\mu_{z}=\frac{\sum_{i} z_{i}}{N}$ is the mean of the vector $\boldsymbol{z}$, $\sigma_{z} = \sqrt{\frac{\sum_{i} (z_{i}-\mu_{z})^{2}}{N}}$. $\alpha$ and $\beta$ are learnable vectors which has the identical shape as $\boldsymbol{z}$. $\alpha$ and $\beta$ will both be fixed during the LLM execution. To apply LayerNorm on the input with a dimension $B\times L\times E$, the LayerNorm will apply to the $B\times L$ vectors, with each has a length of $E$, and $B$ and $L$ denote the batch size and token length, respectively. The intermediate results will then proceed with affine transformation by multiplying with a $E\times 1$ vector $\alpha$ and sum with a $E\times 1$ vector $\beta$.

RMSNorm adopts a more efficient strategy by rescaling the input vectors using only the RMS value of the input vector. Instead of re-centering the data, it directly normalizes the activations based on their RMS value. This approach provides higher computational efficiency compared to LayerNorm. Specifically, RMSNorm can be expressed as:
\begin{equation}
\label{eqn:rmsnorm}
\small
RMSNorm(\boldsymbol{z})=\alpha\frac{\boldsymbol{z}}{r_{z}} + \beta
\end{equation}
where $r_{z} = \sqrt{\frac{\sum_{i} (z_{i})^{2}}{N}}$ is the RMS value of $\boldsymbol{z}$.

\subsection{Efficient Normalization within LLM}
\label{sec:bg:related_work}
Multiple studies have been conducted to accelerate the normalization operation within the transformer architecture. The previous methods including DFX~\cite{hong2022dfx}, ~\cite{jeong2023low} and~\cite{lu2020hardware} modified the way for variance computation, which further improves the parallelism and reduce the processing latency of the layernorm computation. 
In~\cite{wang2023sole}, the intermediate results are dynamically compressed for the variance computation with low precision, leading to a reduction on energy and latency consumption. In~\cite{liu2021hardware}, the parameters of LayerNorm is quantized using 8 bits. However, this will produce a degradation on the accuracy performance.
All the previous mentioned work are proposed for conventional transformer acceleration, which has entirely different scale and data distribution from LLM. These methods often involve rigid quantization and approximate computation, which can be costly for LLMs due to their large size and the high expense of fine-tuning. HAAN selects skipped normalization layers offline with minimal accuracy impact and can be reconfigured to optimize precision and input subsampling for LayerNorm without additional fine-tuning costs.

\section{HAAN Algorithm for Efficient Normalization}
\label{sec:haan_algorithm}
In this section, we first present a statistical study on input distribution of the normalization layers (Section~\ref{sec:input_statistics}). We then demonstrate that variance computation within certain normalization operations can be skipped and propose methods to estimate the variance for these layers (Section~\ref{sec:layernorm_skip}). Finally, we introduce quantization over the operands to facilitate normalization computation (Section~\ref{sec:quantization}).

\subsection{A Study on Input Activation Statistics}
\label{sec:input_statistics}
\begin{figure}
    \centering
    \includegraphics[width=0.45\textwidth]{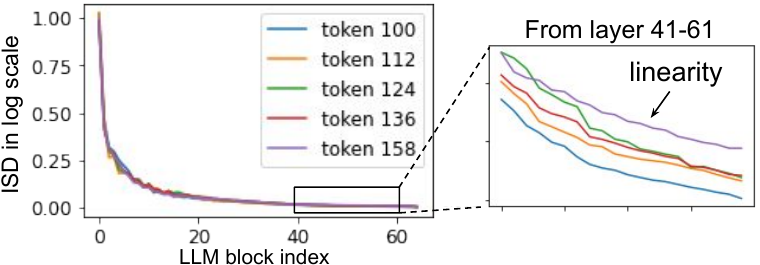}
    \caption{ISD values across different normalization layers within the LLaMA-7B model. Give tokens are chosen randomly. We notice that the ISD values of the later layers reveals a linearity in logarithm domain.}
    \label{fig:rms_trend}
    \vspace{-6mm}
\end{figure}

To motivate our variance skipping algorithm, we examine the trend on the variance of the input activations across all the normalization layers within the LLM. Conducting a statistical analysis in this regard could provide us with additional insights into the correlation between input variance across different layers, thereby enriching our understanding and enabling the generation of more effective optimization strategies. Specifically, rather than variance, we focus on the~\textbf{inverse of the standard deviation (ISD)}, namely $\frac{1}{\sigma}$, which is directly proportional to the normalization results, as shown in equation~\ref{eqn:layernorm} and equation~\ref{eqn:rmsnorm}. More importantly, our profiling results of the normalization operation runtime on GPU reveal that ISD computation accounts for more than 90\% of the overall normalization runtime. Figure~\ref{fig:rms_trend} illustrates the distribution of $\frac{1}{\sigma}$ for each of 64 normalization layers within the LLaMA-7B across different tokens. 
The x-axis represents the layer index, while the y-axis indicates the $\frac{1}{\sigma}$ on a logarithmic scale. We notice that:
\begin{itemize}
\item In general, the ISD values decrease with layer depth, dramatically in the initial layers, then flatten in later layers.
\item The logarithm of ISD values in the deeper layers of the LLM exhibits a pronounced negative linear relationship. 
\end{itemize}

Furthermore, we have replicated the experiment across several other language models, such as GPT-2, OPT, and others, all of which demonstrate a consistent pattern.
The second observation suggests that ISD for the later layers is~\textbf{highly predictable}. This trend in ISD reflects the natural evolution of feature representation within the LLM across various layers: weak token correlation in the initial LLM layer results in pronounced changes in ISD; in the middle layers, as the attention mechanism becomes more intricate, the feature representation stabilizes, leading to linearized changes in ISD; for the deeper layers that are near the end of the LLM outputs, adjustments in feature levels induce significant fluctuations in ISD, a phenomenon associated with the softmax function's capacity to enhance discriminative features~\cite{wang2017normface}. Consequently, we can dynamically skip certain ISD computations and use predicted values, significantly reducing computational costs.

\subsection{Variance Prediction across Layers}
\label{sec:layernorm_skip}
In Section~\ref{sec:input_statistics}, we demonstrate that the ISD exhibits an almost linear relationship when plotted on a logarithmic scale for the latter layers of LLM. This pattern presents a great opportunity for bypassing certain ISD calculations and substituting them with predicted values. Nonetheless, three key questions arise: 
\begin{itemize}
\item Which ISD can be skipped?
\item How can the skipped ISDs be estimated?
\item Will the method exhibit generality across different input samples and datasets?
\end{itemize}

To address the first question, we propose the \textit{ISD skipping algorithm} described in Algorithm~\ref{alg:skiplayer}. A calibration set with $S$ input samples are first fed into the LLM, and their corresponding ISDs are gathered for training the coefficient of the~\textit{ISD predictor}. To pinpoint the ISDs to be skipped, we compute the \textit{Pearson correlation coefficients} across various layer ranges. Subsequently, we identify the range exhibiting the most negative correlation, signifying a significant degree of negative linearity among the ISD values, thus indicating the potential candidates for skipping. Specifically, Algorithm~\ref{alg:skiplayer} iterates over the Pearson correlation coefficients for each pair of ISD values between a pair layer whose gap is larger than $M$, and returning a pairs of layer ranges whose ISD values yield a smallest Pearson correlation coefficient. The ISD computation within this range can be safely skipped and replaced with the predicted values. Moreover, the algorithm also computes the linear gradient of changes in ISD with respect to the layer index gaps utilizing the function \textit{calDecay}. 

To estimate the ISD value, when the layer index $k$ falls within the skip range $(i,j)$, the ISD value of the layer $k$ $(i\leq k \leq j)$ can be estimated using the following equation:
\begin{equation}
log(\text{ISD}_k) = log(\text{ISD}_{i}) + \text{e}_{ij} \times (k-i)
\end{equation}
where i, j represent the start and end layer indices within the range. $\text{ISD}_i$ and $\text{ISD}_i$ are the ISDs of layer i and j. $e_{ij}$ denotes the gradient returned by the $calDecay$ function. Finally, we demonstrate that the ISD predictor exhibits high generalizability across different datasets. Specifically, the ISD predictor trained using the calibration set from one dataset can effectively perform well on other downstream tasks. The evaluation results are detailed in Section~\ref{sec:accuracy_evaluation}.

\begin{figure}[tp]
    \centering
    \small
    \begin{algorithm}[H]
        \setstretch{1}
        \caption{ISD Skipping Algorithm}\label{alg:skiplayer}
        \begin{algorithmic}[1]
        \Require Calibration set \textbf{S}, LLM model with $L$ layers, $F_{l}(.)$, skip range $M$
        \Ensure Skipping range $(i_{f}, j_{f})$, coefficient $e$
        \State Initialize $ISDLists \gets \emptyset$, $minCor \gets 1$
        \ForAll{$s \in S$}
            \ForAll{$l \in L$}
                \State Compute $F_{l}(s)$, record $\log(v_{l})$ in $ISDLists$
            \EndFor
        \EndFor
            
        \For{$i \gets 1$ to $L - M$}
            \State $j \gets i + M$
            \If{$Pearson(ISDLists[i:j], [i:j]) < minCor$}
                \State $minCor \gets Pearson(ISDLists[i:j], [i:j])$
                \State $(i_{f}, j_{f}) \gets (i, j)$
                \State $e \gets \text{calDecay}(ISDLists[i:j])$
            \EndIf
        \EndFor
        
        \State \Return $(i_{f}, j_{f})$, $e$
        \end{algorithmic}
    \end{algorithm}
    \vspace{-5mm}
\end{figure}

\subsection{Efficient Normalization with Subsampling and Quantization}
\label{sec:quantization}
In addition to the ISD skipping method outlined in Section~\ref{sec:layernorm_skip}, two additional approaches are introduced in this section to improve the efficiency of normalization. First, for the remaining ISD values that cannot be skipped, their values can be estimated by using a subsampled version of the input:
\begin{equation}
\small
\text{ISD}_{sub} = \frac{1}{\sqrt{\frac{1}{N_{sub}} \sum_{a=1}^{N_{sub}} z_a^2}}
\end{equation}
where $\text{ISD}_{sub}$ represents the estimated version of the ISD using the subsampled inputs, $N_{sub}$ denotes the number of input samples used for ISD estimation, and $z_a$ signifies the $a$-th input $\boldsymbol{z}$. Empirically, a minimal $N_{sub}$ is chosen to maintain a negligible impact on perplexity (PPL). It is important to note that the optimal value of $N_{sub}$ varies between different LLM models, as discussed in Section~\ref{sec:accuracy_evaluation}. To implement the subsampling operation on the input, we simply truncate the first $N_{sub}$ elements within the input.

Moreover, subsampling can also aid in streamlining mean computation within LayerNorm. As demonstrated in Section~\ref{sec:accuracy_evaluation}, a substantial portion of the input can be discarded to estimate of both mean and ISD with negligible impact on LLM accuracy. In addition to subsampling, proper quantization of operands during normalization is also applied to reduce implementation costs.

Our normalization hardware accelerator, as detailed in Section~\ref{sec:haan_hardware}, can effectively accommodate all of the optimization techniques described above, including ISD skipping, subsampling, and quantization~\cite{jacob2018quantization}. Additionally, it is reconfigurable in the sense that it can also execute standard normalization computation without employing any optimization techniques, this guarantees the robustness of the accelerator performance.

\section{HAAN Hardware Accelerator}
\label{sec:haan_hardware}

\begin{figure}
    \centering
    \includegraphics[width=0.45\textwidth]{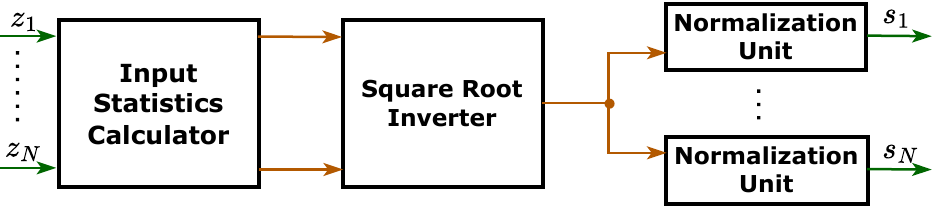}
    \caption{Data path of HAAN where $N$ denotes the size of embedding dimensions. In this figure, the floating-point and fixed-point data paths are highlighted in green and orange, respectively. The control signals are highlighted in black.
    }  
    \label{fig:hardware-overview}
    \vspace{-4mm}
\end{figure}
This section details the hardware design for accelerating the normalization operation. The overall architecture of the HAAN accelerator is shown in Figure~\ref{fig:hardware-overview}. The system consists of three main elements: an \textit{input statistic calculator}, a \textit{square root inverter}, and \textit{normalization units}. The input statistic calculator computes the mean and variance of the input tensor. The square root inverter uses these intermediate results to produce the ISD, and the normalization units generate the normalized input and perform the affine transformation. 
Our hardware design is highly configurable, supporting input and output in either fixed-point or floating-point (FP16 or FP32) formats while maintaining intermediate computational results in fixed-point representation. To reduce computational costs, the input statistics calculator and normalization units allow statistics to be collected from a subset of the original input. Combined with the techniques in Section~\ref{sec:quantization}, this approach significantly reduces hardware costs with minimal impact on accuracy.

\subsection{Input Statistics Calculator}
\label{sec:input_statistic_compute}
\begin{figure}
    \centering
    \includegraphics[width=0.45\textwidth]{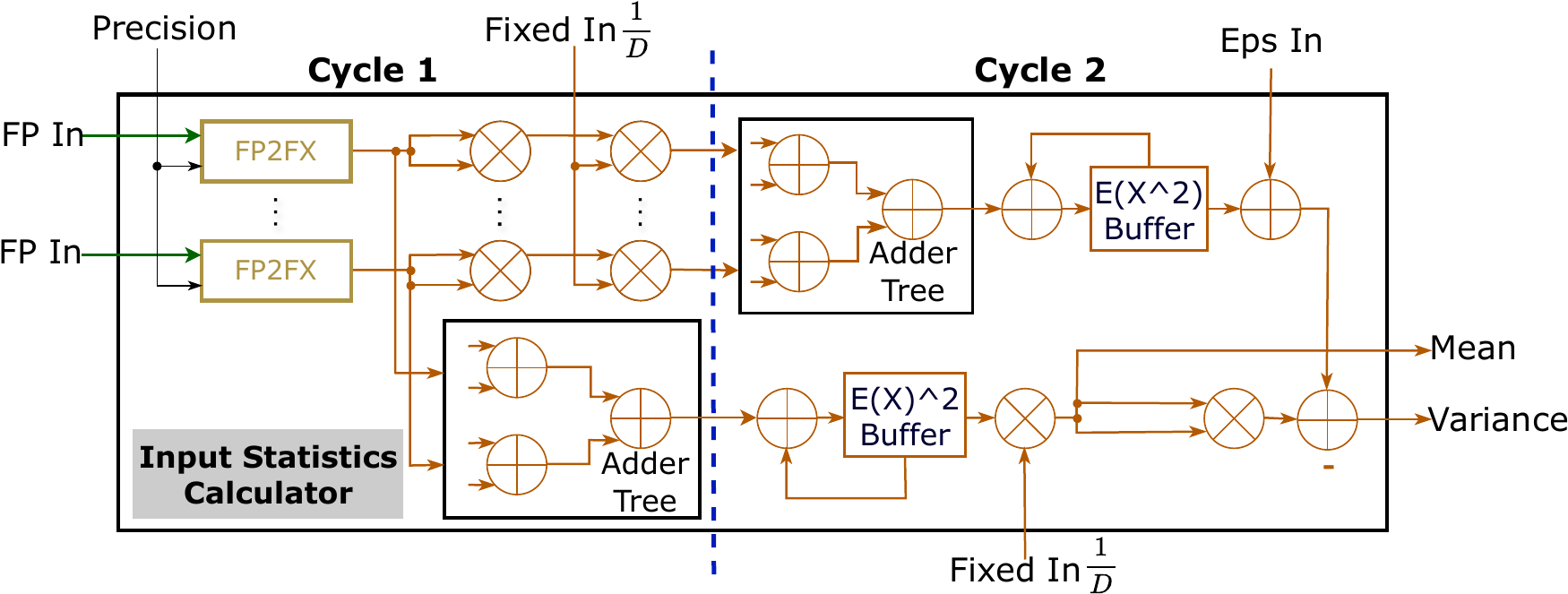}
    \caption{Hardware design for Input Statistics Calculator.}
    \label{fig:mean-var}
    \vspace{-6mm}
\end{figure}

The purpose of the Input statistic calculator is to generate the mean and variance for a D-dimensional vector $\boldsymbol{z}$, the hardware design is shown in Figure~\ref{fig:mean-var}. Specifically, to reduce the processing latency, the variance of $\boldsymbol{z}$ can be expressed with the following equation:
\begin{equation}
    \label{var-eq}
    \small
    Var(\boldsymbol{z})=E(\boldsymbol{z}^2)-(E(\boldsymbol{z}))^2=\sum\limits_{i} \frac{z_i^2}{N} - (\frac{1}{N}\sum\limits_{i} z_i)^2
\end{equation}
where $N$ is the input dimension of $\boldsymbol{z}$. The input statistics calculator will compute $\sum\limits_{i} \frac{z_i^2}{N}$ and $ (\frac{1}{N}\sum\limits_{i} z_i)^2$ in parallel. 
During execution, the inputs \(\boldsymbol{z}\) in FP format are first converted to fixed-point format using FP2FX units. If the inputs are already in fixed-point format (INT8), the FP2FX units will bypass the conversion. Two multiplication units then process each element \(z_{i}\) of \(\boldsymbol{z}\) to generate \(\frac{z_i^2}{N}\). Since \(N\) is predetermined, \(\frac{1}{N}\) can be precomputed and stored in memory. If \(N\) is a power of 2, division is implemented with a shift operation. An adder tree is then used to calculate \(\sum_{i} \frac{z_i^{2}}{N}\).
Concurrently, another adder tree concurrently computes \(\sum_{i} z_i\) and the mean. The multiplication unit then calculates \((\frac{1}{N}\sum_{i}z_{i})^2\), followed by a subtractor to produce the variance of \(\boldsymbol{z}\). Due to the high dimensionality of \(\boldsymbol{z}\) in LLMs, the mean and variance computation occurs over multiple passes, with interim results stored in local registers. For RMSNorm, the mean is not required and can be omitted. Using the subsampling techniques from Section~\ref{sec:haan_algorithm} reduces passes and cycles, further decreasing processing latency.

\subsection{Square Root Inverter}
\label{sec:inverted_square_root_unit} 

The variance produced from the input statistics calculator will be further processed by the square root inverter, which will accept FP number $x$ and produce the output $y = \frac{1}{\sqrt{x}}$. Before the processing start, the FX2FP units will first convert the variance into FP format. Denote the $E_{x}$ and $M_{x}$ and exponent and mantissa fields of $x$. For positive x, 
we have $x=2^{E_x-Q}(1+M_x/2^{L})$, where Q is the bias and L is the bit length of mantissa field. Then $\log_2 x = E_x-Q + \log_2(1+M_x/2^{L})\approx E_x-Q + M_x/2^{L} +\sigma_x=\frac{2^{L}(E_x-Q) + M_x}{2^{L}} + \sigma_x$. Where $\sigma_{x}$ can be approximated as a constant $0.450465$ when $M_x/2^{L}\in[0,1]$ with negligible error~\cite{lomont2003fast}.
To compute $y = \frac{1}{\sqrt{x}}$, we first calculate the $\log_2(y)$, which can be derived as follows: 
{\small
\begin{align}
    \scriptsize
    \log_2 (y) &= -\frac{1}{2} \log_2 (x) \label{log} \\
    \frac{2^{L}(E_y-Q) + M_y}{2^{L}} + \sigma_y &= -\frac{1}{2}(\frac{2^{L}(E_x-Q) + M_x}{2^{L}} + \sigma_x) 
\end{align}
\begin{align}
    \small
    \begin{split}
        M_y+2^{L}E_y &=  -\frac{1}{2}(M_x +2^{L}E_x) + 2^L(\frac{3}{2}Q - (\sigma_y + \frac{1}{2}\sigma_x)) \\ 
    & \approx -\frac{1}{2}(M_x +2^{L}E_x) + \frac{3}{2}2^{23} (127-0.450465)\\ 
    & \approx 0\times5f3759df-\frac{1}{2}(M_x+2^{L}E_x) \label{fast-invsqrt}
    \end{split}
\end{align}
}
where $M_y+2^{L}E_y$ is the bit representation of the FP format of y. 
\begin{figure}
    \centering
    \includegraphics[width=0.45\textwidth]{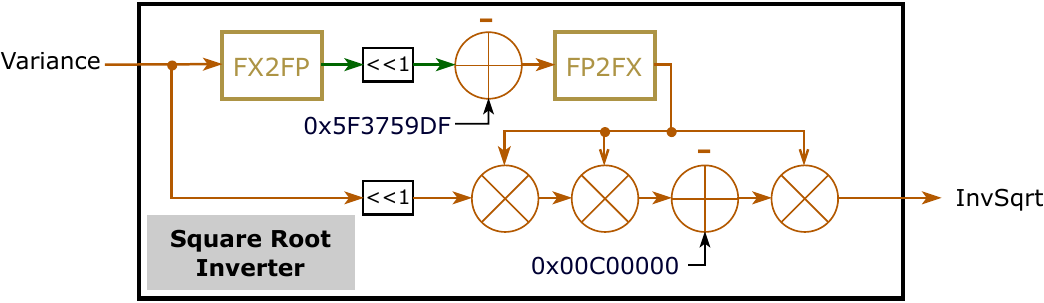}
    \caption{Hardware design of Square Root Inverter. In the figure $0x00C00000$ equals $1.5$ in fixed-point format.}
    \label{fig:invsqrt}
    \vspace{-6mm}
\end{figure}
To enhance the accuracy of the inverted square root operation, we first convert the y from the FP format to fixed-point format, and utilize Newton's method to refine the initial approximation obtained from equation~\ref{fast-invsqrt}. Let $f(y) = \frac{1}{y^{2}}-x$, and denote $y_{0}$ as the initial value returned from equation~\ref{fast-invsqrt}. Following Newton's method, the subsequent value can be computed as follows:
\begin{equation}
     \label{eqn:newton-iter}
     \small
     y_{1} = y_0 - \frac{f(y_0)}{f^{'}(y_0)} = y_0- \frac{1/y_0^2 - x}{-2/y_0^3}=y_0 (1.5-xy_0^2)
\end{equation}
Based on the evaluation results, we observe that a single iteration is adequate to achieve accurate results. The architecture of Square Root Inverter is depicted in Figure~\ref{fig:invsqrt}.

To support the layer skipping methods from Section~\ref{sec:layernorm_skip}, we design a custom unit to calculate predicted ISD using previous statistics. It employs the coefficient \(e\) of the ISD predictor and ISD values from early layers, leveraging the Xilinx Floating-point IP Core for linear prediction in the logarithm domain. The prediction is sent to the normalization unit. The ISD predictor is a scalar processor with minimal hardware cost.

\subsection{Normalization Unit and Memory Layout}
\label{sec:normalization_unit}
The normalization unit receives external inputs and outputs from the input statistics calculator and square root inverter to generate normalized results. For certain normalization layers, the ISD skipping algorithm in Algorithm~\ref{alg:skiplayer} allows direct prediction of ISD values, bypassing the square root inverter. The normalized input then undergoes affine transformation by multiplying by \(\alpha\) and adding \(\beta\). When quantization is enabled, outputs remain in fixed-point format, skipping conversion in the FX2FP units. The input statistics calculator, square root inverter, and normalization unit operate in a pipelined manner across multiple input samples.

\begin{figure}
    \centering
    \includegraphics[width=0.4\textwidth]{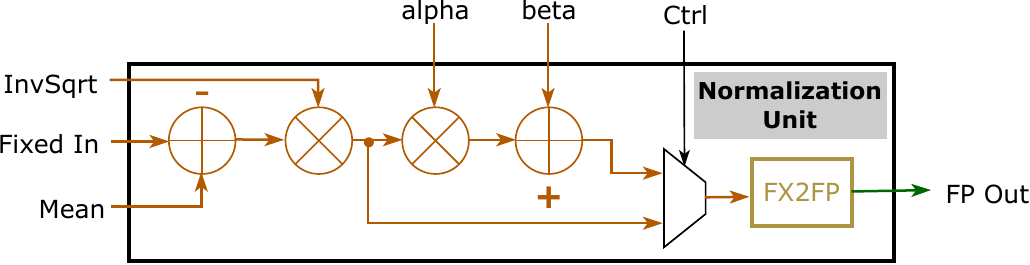}
    \caption{Hardware design of the Normalization Unit}
    \label{fig:norm}
\vspace{-5mm}
\end{figure}

\begin{figure}
    \centering
    \includegraphics[width=0.34\textwidth]{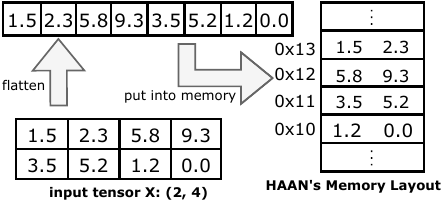}
    \caption{HAAN memory layout for one sample $\boldsymbol{x}$ with a dimension of $2\times 4$.}
    \label{fig:memory}
    \vspace{-6mm}
\end{figure}

To support HAAN execution, we design an efficient storage format shown in Figure~\ref{fig:memory}. The input tensor is flattened into a vector, with each memory entry containing a chunk equal to the input bandwidth of the HAAN accelerator. This allows the accelerator to access and process one memory entry per cycle. Figure~\ref{fig:memory} shows storing an input sample with dimensions \(2 \times 4\), where the accelerator processes a vector of length two per cycle. In subsampling mode, only the initial portion of memory entries is accessed for computing input statistics.

\section{EXPERIMENTAL RESULTS}
\label{sec:evaluation}
In this section, we first evaluate the HAAN algorithm in Section~\ref{sec:accuracy_evaluation}. Next, we present the hardware system evaluation in Section~\ref{sec:hardware_evaluation}.
\subsection{Accuracy Evaluation}
\label{sec:accuracy_evaluation}

\begin{table}[]
\vspace{-4mm}
\captionsetup{font=small}
\tiny
\centering
\resizebox{0.43\textwidth}{!}{%
\begin{tabular}{ c|c|c c c c c }
\hline
model &
  method &
  WG &
  PQ &
  HS &
  A-e &
  A-c \\ \hline
 &
  Original &
  {\color[HTML]{000000} 0.7017} &
  {\color[HTML]{000000} 0.7867} &
  {\color[HTML]{000000} 0.5694} &
  {\color[HTML]{000000} 0.7517} &
  {\color[HTML]{000000} 0.4198} \\ \cline{2-7} 
\multirow{-2}{*}{LLaMA-7B} &
  HAAN &
  {\color[HTML]{000000} 0.7016} &
  {\color[HTML]{000000} 0.7818} &
  {\color[HTML]{000000} 0.5696} &
  {\color[HTML]{000000} 0.7567} &
  {\color[HTML]{000000} 0.4163} \\ \hline
 &
  Original &
  {\color[HTML]{000000} 0.6093} &
  {\color[HTML]{000000} 0.7367} &
  {\color[HTML]{000000} 0.4581} &
  {\color[HTML]{000000} 0.6073} &
  {\color[HTML]{000000} 0.2696} \\ \cline{2-7} 
\multirow{-2}{*}{OPT-2.7B} &
  HAAN &
  {\color[HTML]{000000} 0.6085} &
  {\color[HTML]{000000} 0.7318} &
  {\color[HTML]{000000} 0.4582} &
  {\color[HTML]{000000} 0.5997} &
  {\color[HTML]{000000} 0.2713} \\ \hline
 &
  Original &
  {\color[HTML]{000000} 0.5833} &
  {\color[HTML]{000000} 0.7084} &
  {\color[HTML]{000000} 0.4004} &
  {\color[HTML]{000000} 0.5829} &
  {\color[HTML]{000000} 0.2500} \\ \cline{2-7} 
\multirow{-2}{*}{GPT2-1.5B} &
  HAAN &
  {\color[HTML]{000000} 0.5801} &
  {\color[HTML]{000000} 0.7065} &
  {\color[HTML]{000000} 0.3997} &
  {\color[HTML]{000000} 0.5779} &
  {\color[HTML]{000000} 0.2554} \\ \hline
\end{tabular}%
}
\caption{Accuracy evaluation of HAAN on PIQA (PQ), WinoGrande (WG), HellaSwag (HS), Arc-Easy (A-e), Arc-Challenge (A-c) tasks.}
\label{tab:my-eval}
\vspace{-3mm}
\end{table}

\subsubsection{Accuracy evaluation over different LLMs and tasks}

We run simulations of the HAAN algorithm, as detailed in Section~\ref{sec:haan_algorithm}, using the Huggingface library. All pretrained LLMs are downloaded from Huggingface's official repository. To assess LLMs' accuracy, we consider five tasks: PIQA~\cite{bisk2020piqa}, WinoGrande~\cite{sakaguchi2021winogrande}, HellaSwag~\cite{zellers2019hellaswag}, and Arc (Easy and Challenge)~\cite{clark2018think}. We employ the evaluation~\cite{eval-harness} with default parameters for our experiments on NVIDIA RTX 3090Ti GPUs.

For the ISD skipping algorithm outlined in Algorithm~\ref{alg:skiplayer}, we random select 100 samples from the Wikitext dataset and use it as the calibration set. Subsequently, we execute the ISD algorithm across various LLMs and record the skip range for each. This recorded range will be applied to streamline the normalization computation across different downstream tasks. Specifically, for the LLaMA-7B model, we utilize the first $N_{sub}=256$ input sample with a skip range of $(i_f,j_f) = (50, 60)$ to estimate the ISD. Furthermore, we apply INT8 quantization over the input. For OPT-2.7B model, we utilize the first $N_{sub}=1280$, with the skip range adjusted to $(i_f,j_f) = (55, 62)$, and FP16 precision was employed. The GPT2-1.5B model is configured with a $N_{sub}=800$ and a skip range of (85, 92), also utilizing FP16 precision. 

Each LLM's performance is compared with the original setting in Table~\ref{tab:my-eval}, where the original denotes the FP32 LLM without ISD skipping and subsampling. The results show HAAN achieves nearly the same accuracies ($<1\%$ degradation) as the original LLM across three models and five tasks. This demonstrates that HAAN is highly reliable.

\subsubsection{Ablation studies}
In this section, we investigate the impact of skip range $(i_f,j_f)$, subsampling length ($N_{sub}$) and data format over the LLM accuracy. We choose LLaMA to study and evaluate its accuracy with different configurations. The evaluation results are shown in Table~\ref{tab:self-eval}. We have the following observation: first, $N_{sub}$ has a significant impact over the accuracy. Particularly, $N_{sub}=256$ achieves a great balance between efficiency and precision. 
Furthermore, in terms of data precision, INT8, FP16 and FP32 all achieve comparable results. Finally, the selection of the skip range is critical: skipping ISD computation in the early part of normalization layers of the model severely harm the model accuracy, leading to a failure on A-c task. 
Skipping in the middle also causes a notable accuracy drop.
\begin{table}[]
\captionsetup{font=small}
\tiny
\centering
\resizebox{0.45\textwidth}{!}{%
\begin{tabular}{c|c|ccccc}
\hline
{\color[HTML]{000000} Method} &
  {\color[HTML]{000000} Config} &
  {\color[HTML]{000000} WG} &
  {\color[HTML]{000000} PQ} &
  {\color[HTML]{000000} HS} &
  {\color[HTML]{000000} A-e} &
  {\color[HTML]{000000} A-c} \\ \hline
{\color[HTML]{000000} } &
  {\color[HTML]{000000} 128} &
  \multicolumn{1}{r}{{\color[HTML]{000000} 0.5722}} &
  \multicolumn{1}{r}{{\color[HTML]{000000} 0.6654}} &
  \multicolumn{1}{r}{{\color[HTML]{000000} 0.4067}} &
  \multicolumn{1}{r}{{\color[HTML]{000000} 0.4520}} &
  \multicolumn{1}{r}{{\color[HTML]{000000} 0.2432}} \\ \cline{2-7} 
{\color[HTML]{000000} } &
  {\color[HTML]{000000} 256} &
  {\color[HTML]{000000} 0.7016} &
  {\color[HTML]{000000} 0.7818} &
  {\color[HTML]{000000} 0.5696} &
  {\color[HTML]{000000} 0.7567} &
  {\color[HTML]{000000} 0.4163} \\ \cline{2-7} 
\multirow{-3}{*}{{\color[HTML]{000000} Subsample length}} &
  {\color[HTML]{000000} 512} &
  {\color[HTML]{000000} 0.7015} &
  {\color[HTML]{000000} 0.7828} &
  {\color[HTML]{000000} 0.5691} &
  {\color[HTML]{000000} 0.7513} &
  {\color[HTML]{000000} 0.4168} \\ \hline
{\color[HTML]{000000} } &
  {\color[HTML]{000000} INT8} &
  {\color[HTML]{000000} 0.7016} &
  {\color[HTML]{000000} 0.7818} &
  {\color[HTML]{000000} 0.5696} &
  {\color[HTML]{000000} 0.7567} &
  {\color[HTML]{000000} 0.4163} \\ \cline{2-7} 
{\color[HTML]{000000} } &
  {\color[HTML]{000000} FP16} &
  {\color[HTML]{000000} 0.7016} &
  {\color[HTML]{000000} 0.7826} &
  {\color[HTML]{000000} 0.5691} &
  {\color[HTML]{000000} 0.7545} &
  {\color[HTML]{000000} 0.3963} \\ \cline{2-7} 
\multirow{-3}{*}{{\color[HTML]{000000} Data format}} &
  {\color[HTML]{000000} FP32} &
  {\color[HTML]{000000} 0.7017} &
  {\color[HTML]{000000} 0.7862} &
  {\color[HTML]{000000} 0.5691} &
  {\color[HTML]{000000} 0.7511} &
  {\color[HTML]{000000} 0.4198} \\ \hline
{\color[HTML]{000000} } &
  {\color[HTML]{000000} $(10, 20)$} &
  {\color[HTML]{000000} 0.5018} &
  {\color[HTML]{000000} 0.5818} &
  {\color[HTML]{000000} 0.3496} &
  {\color[HTML]{000000} 0.5032} &
  {\color[HTML]{000000} 0.2512} \\ \cline{2-7} 
{\color[HTML]{000000} } &
  {\color[HTML]{000000} $(30, 40)$} &
  {\color[HTML]{000000} 0.6218} &
  {\color[HTML]{000000} 0.7018} &
  {\color[HTML]{000000} 0.4896} &
  {\color[HTML]{000000} 0.6767} &
  {\color[HTML]{000000} 0.2675} \\ \cline{2-7} 
\multirow{-3}{*}{{\color[HTML]{000000} Skip range}} &
  {\color[HTML]{000000} $(50, 60)$} &
  {\color[HTML]{000000} 0.7016} &
  {\color[HTML]{000000} 0.7818} &
  {\color[HTML]{000000} 0.5696} &
  {\color[HTML]{000000} 0.7567} &
  {\color[HTML]{000000} 0.4163} \\ \hline
\end{tabular}%
}
\caption{LLaMA-7B accuracy across different configurations.}
\label{tab:self-eval}
\vspace{-4mm}
\end{table}

\begin{figure}
    \centering
    \includegraphics[width=0.46\textwidth]{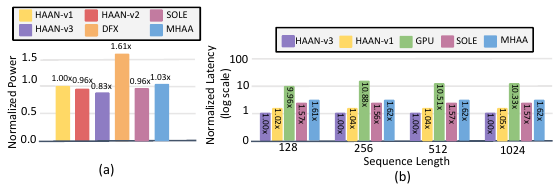}
    \caption{(a) Normalized power comparison. (b) Normalized latency of HAAN compared to GPU on OPT-2.7B. The results of \textit{HAAN-v2} are excluded from our evaluation since its configuration is incompatible with GPT2 model.}
    \label{fig:power-comparison}
\vspace{-7mm}
\end{figure}
\subsection{Hardware Evaluation} 
\label{sec:hardware_evaluation}
In this section, we describe the hardware performance of the HAAN accelerator. We implement HAAN on the Xilinx Alveo U280 board and synthesize our design using the Xilinx Vivado Suite. The HAAN accelerator operates at a 100MHz clock frequency.
The HAAN accelerator is highly reconfigurable, supporting different input data widths for both the input statistics calculator and the normalization unit, denoted by \(p_d\) and \(p_n\) respectively. When subsampling is enabled, the parallelism of the input statistics calculator can be reduced without increasing processing latency, freeing up hardware resources (e.g., DSP) for more normalization units with more pipeline levels. The HAAN accelerator also supports various input data formats, including FP32, FP16, and INT8.

\subsubsection{Self-Evaluation Under Different Configurations}
We provide a comprehensive evaluation by varying the data formats, token length and subsampling rate over the normalization layer input.
We report the corresponding hardware resource usage and power consumption. The corresponding resource and power costs under different settings are reported in Table~\ref{tab:haan_selfstudy}. We notice that performing the normalization in FP32 consumes $1.29\times$ more power than FP16 processing on average, and the power consumption increases moderately as the input sequence length grows. Overall, the normalization in INT8 achieves the least hardware and power consumption.

\subsubsection{Performance comparison with other systems}

We assess the hardware performance of the HAAN accelerator against other accelerators for normalization operation over GPT-2 and OPT.
Most previous accelerators for LayerNorm~\cite{wang2023sole}~\cite{10326397}~\cite{9976354} are either designed and evaluated with ASIC or not revealed the detailed architecture design~\cite{10288182}~\cite{lu2020hardwareacceleratormultiheadattention}. To ensure a fair comparison, we compare HAAN accelerator with DFX~\cite{hong2022dfx} by extracting the latency of LayerNorm from the overall latency reported in their study. We also reproduce SOLE~\cite{wang2023sole} and MHAA~\cite{lu2020hardwareacceleratormultiheadattention} while aligning with HAAN's settings to ensure a fair comparison.

With the ISD skipping algorithm, ten normalization layers in GPT-2 can be skipped, and the input can be subsampled by half of its length. This modification does not cause noticeable degradation ($\leq 0.5\%$) in LLM accuracy on five downstream tasks described in Section~\ref{sec:accuracy_evaluation}. For a comprehensive evaluation, we include two configurations for our HAAN accelerator, denoted as \textit{HAAN-v1} and \textit{HAAN-v2}. Both HAAN-v1 and HAAN-v2 adopt a single pipeline and accept normalization input in FP16. The $(p_d, p_n)$ are set to (128, 128) and (80, 160) for HAAN-v1 and HAAN-v2, respectively.

\begin{table}
    \tiny
    \captionsetup{font=small}
    \centering
    \resizebox{0.48\textwidth}{!}{%
        \begin{tabular}{|c|c|c|c|c|c|}
            \hline
            \multirow{2}{*}{Input Format} & \multirow{2}{*}{$(p_d,p_n)$}  & \multicolumn{3}{c|}{FPGA resource consumption}  & \multirow{2}{*}{Power (W)}\\ \cline{3-5}
                &  & LUT & FF & DSP & \\ \hline
             \multirow{2}{*}{FP32}  & (128, 128) & 84K/4.9\% & 17K/0.5\% & 1536/12.5\% & 6.362\\ \cline{2-6}
            & (32, 128) & 99K/5.7\% & 21K/0.6\% & 1036/8.4\% & 6.136\\ \hline
            \multirow{2}{*}{FP16}  & (128, 128) & 55K/3.2\% & 11K/0.4\% & 1536/12.5\% & 4.868\\ \cline{2-6}
            & (32, 128) & 76K/4.5\% & 15K/0.4\% & 1036/8.4\% & 4.790 \\ \hline
            \multirow{2}{*}{INT8}  & (256, 256) & 58K/3.4\% & 21K/0.6\% & 1536/12.5\% & 3.458\\ \cline{2-6}
            & (32, 512) & 86K/5.0\% & 25K/0.7\% & 1025/8.3\% & 6.382 \\ \hline
        \end{tabular}
    }
    \caption{Hardware cost of HAAN accelerator with FP32/FP16/INT8 inputs. The left/right numbers represent absolute values and percentage of overall resources. Power was measured as the average at input lengths of 16, 128, and 256.}
    \label{tab:haan_selfstudy}
    \vspace{-5mm}
\end{table}

\begin{figure}
    \centering
    \includegraphics[width=0.45\textwidth]{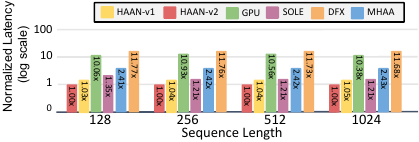}
    \caption{Normalized latency of HAAN compared to DFX and GPU on GPT2-1.5B. } 
    \label{fig:latency-gpt}
\vspace{-7mm}
\end{figure}

Figure~\ref{fig:power-comparison} (a) and Figure~\ref{fig:latency-gpt} compare processing latency and power consumption for normalization layers in GPT-2. HAAN-v1 and HAAN-v2 show an average latency reduction of $11.7\times$, $10.5\times$, $1.25\times$ and $2.42\times$ compared to DFX, GPU, SOLE and MHAA, respectively, across different input sequence lengths. Additionally, HAAN-v1 and HAAN-v2 achieve average power usage reductions of $61\%$ and $64\%$ compared to DFX while using slightly less power than SOLE and MHAA. 
HAAN's superior performance primarily stems from its ISD skipping and subsampling mechanism, leading to significant computational cost reduction. Additionally, HAAN optimizes the latency of each individual blocks by dynamically configuring the precision, enabling it to support deep pipelining for enhanced throughput. Also, HAAN utilizes inter-sample parallelism and pipelining and by setting particular $p_d, p_n$, the time of the different stages of the pipeline is evenly distributed, so that we can maximize the utilization rate of hardware units and lower latency when input size becomes huge. 

Figure~\ref{fig:power-comparison} (b) illustrates the comparative analysis of processing latency of the normalization layers within OPT-2.7B, where 7 out of 65 ISD operations can be skipped and the input can be further truncated with a length of 1280. Here we introduce \textit{HAAN-v3}, another HAAN configuration with a single pipeline, FP16 input and $(p_d,p_n)$ set to (64,128). We also observe that HAAN-v1 and HAAN-v3 achieve a notable reduction in processing latency compared to the baseline hardware platforms.

To evaluate end-to-end performance, we use the experimental settings and results from previous work~\cite{Chen_2024}. We test the GPT-2 model with 355M parameters and 24 layers, experimenting with input lengths of 128, 256, and 512. Our results show that incorporating HAAN will enable an average end-to-end speedup of approximately $1.11\times$ across different input lengths.

\section{Conclusion}
This study introduces HAAN, a holistic normalization acceleration approach for LLMs. HAAN exploits the correlation in normalization statistics among adjacent layers to bypass normalization computation by estimating statistics from preceding layers. This evaluation results demonstrate that HAAN can achieve great improvement over other baselines algorithms. 

\bibliography{ref}
\bibliographystyle{IEEEtran}

\end{document}